\documentclass[12pt]{article}
\usepackage{epsfig}
\begin{document}

\title{Selected Experimental Highlights from Nucleus-Nucleus Collisions at RHIC}         
\author{Huan Zhong Huang \\
	Department of Physics and Astronomy \\
	University of California \\
	Los Angeles, CA 90095-1547, USA}        
\date{}          
\maketitle

\abstract{Nucleus-nucleus collisions at RHIC produce high temperature and high energy density matter which exhibits partonic degrees of freedom. We will discuss measurements of nuclear modification factors for light hadrons and non-photonic electrons from heavy quark decays, which reflect the flavor dependence of energy loss of high momentum partons traversing the dense QCD medium. The hadronization of bulk partonic matter exhibits collectivity in effective partonic degrees of freedom. Nuclear collisions at RHIC provide an intriguing environment, where many constituent quark ingredients are readily available for possible formation of exotic particles through quark coalescences or recombinations.}

\section{Introduction}       
Lattice Quantum ChromoDynamics (QCD) predicted that at high temperature and/or high baryon density quarks and gluons are 
deconfined and a new state of matter, the Quark-Gluon Plasma (QGP) is formed. The nature of the transition from hadrons to the QGP has not been established. Theoretical calculations indicate that there may be a critical point at finite baryon density, and at low baryon density the transition may be a smooth cross over while 
at higher baryon density a first-order phase transition may take place. More details may be found in~\cite{karsch} and references therein.

High energy nucleus-nucleus collisions have been used to study QCD at the extreme conditions of temperate and density and to search for the new state of matter. Experimental results at RHIC have demonstrated that matter of extreme high energy density has been produced in central Au+Au collisions. High momentum partons from hard scatterings have been used to study the energy loss of the partons when travesing the dense matter. In this brief report we will discuss the flavor dependence of the parton energy loss in dense medium, characteristics of hadronization for bulk partonic matter and possible formation of exotic particles from nuclear collisions at RHIC.  
  
\section{Partonic Energy Loss in Dense QCD Medium}
High energy heavy ion collisions at RHIC made it possible to realize an idea first proposed by Bjorken~\cite{bj} and further developed by X.N. Wang and M. Gyullasy~\cite{wg} to study partons scattering off high temperature dense matter. Nuclear modification factors, 
\[ R_{AA} = \frac{[\frac{dn}{dp_{T}}]_{_{AA}}/N_{\rm{coll}}}{[\frac{dn}{dp_{T}}]_{_{pp}}} \hspace{0.25in}{\rm{and}}\hspace{0.25in}
 R_{CP} = \frac{[\frac{dn}{dp_{T}}]_{_{Cen}}/N^{\rm{Cen}}_{\rm{coll}}}{[\frac{dn}{dp_{T}}]_{_{Per}}/N^{\rm{Per}}_{\rm{coll}}}, \]
have been defined to quantitatively describe the change in the $p_{T}$ spectra due to energy loss of high momentum partons in the dense medium, where $N_{\rm{coll}}$ denotes for the number of binary inelastic nucleon-nucleon collisions, $\rm{Cen}$ 
and $\rm{Per}$ for central and peripheral nucleus-nucleus (AA) collisions. These factors would be unity if nucleus-nucleus collisions are a superposition of independent nucleon-nucleon collisions. The experimental observations of a strong suppression of high transverse momentum ($p_{T}$) particles ($R_{AA}$ or $R_{CP}\sim 0.2$ $<<$ 1) in central Au+Au collisions have been one of the most exciting discoveries at RHIC, which established that the presence of dense matter created in central Au+Au collisions significantly changes the high $p_T$ distribution of hadrons produced from partons traversing the dense medium. 

Recent experimental studies of high $p_T$ physics at RHIC focus on questions of 1) whether the suppression of high $p_T$ particles in central Au+Au collisions is mainly due to partonic, not hadronic, energy loss; 2) whether there is experimental effects due to quark versus gluon energy loss difference; and 3) whether there is a flavor dependence in partonic energy loss, in particular, difference between light and heavy quarks. 

Figure~1 shows the nuclear modification factors $R_{AA}$ of neutral $\pi^0$ and $\eta$ from PHENIX~\cite{phe-pi}
and $R_{CP}$ of protons and charged pions from STAR~\cite{star-ppi}. The boxes on STAR data are systematic errors and the PHENIX errors contain point-by-point statistical and systematic errors.  These data have extended the measurements of nuclear modification factors for identified particles to much higher $p_T$ and confirm the previous observation that there is a $p_T$ scale around 6 GeV/c that distinct particles produced by fragmentation processes from those at the 
intermediate $p_T$ of 2-6 GeV/c where multi-parton dynamics are important~\cite{star-L}. Note that the definitions of collision centrality and nuclear modification factors are slightly different for STAR and PHENIX data. 

\begin{figure}[htb]
   \centering
   \epsfxsize=4.0in
   \epsfysize=2.5in
   \leavevmode
   \epsffile{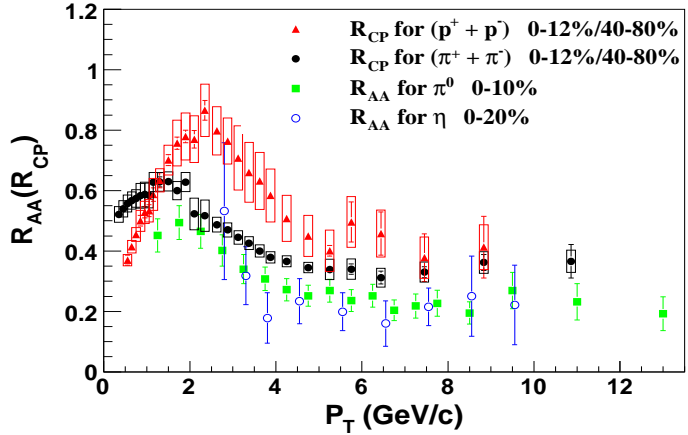} 
\parbox{5.5in}{\caption[prot]{Nuclear modification factors $R_{AA}$ for $\pi^0$ and $\eta$ from PHENIX and $R_{CP}$ for $\pi^{+}+\pi^{-}$ and proton ($p+\overline{p}$) from STAR. The differences in $R_{AA}$ between $\pi^0$ and $\eta$, in $R_{CP}$ between $\pi$ and proton disappear above $p_{T}$ of 6 GeV/c indicative of parton energy loss before the hadron formation.}}
\label{prot}
\end{figure}   

These measurements of nuclear modification factors indicate no significant difference between $\pi^0$ and $\eta$, between charged $\pi$ and protons (anti-protons) in the region of $p_T$ greater than 6 GeV/c. Therefore, the energy loss processes must have taken place before the formation of these particles, which have very different hadronic scattering cross sections if energy loss from hadronic re-scattering were to dominate. These data have experimentally established that the energy loss dynamics must be due to partons, which have long been assumed in theoretical calculations of jet 
quenching~\cite{xnwang,vitev,wiedemann}. More importantly there is no significant difference between baryons and mesons as shown by the STAR data. Gluon and quark jet fragmentations tend to contribute differently to baryons and mesons and generally anti-baryons at high $p_T$ tend to originate more from gluon fragmentations~\cite{delph}. The large difference expected theoretically for gluon and quark energy loss in QCD medium does not manifest in the high $p_T$ suppression of baryons and mesons from central Au+Au collisions. Recently Liu, Ko and Zhang calculated the quark-gluon jet conversion rate in the QGP using the lowest order QCD and they found that a much larger conversion rate is needed in order to explain the STAR measurement~\cite{cmko}. Questions remain open whether and by what dynamics high $p_T$ gluons would immediately convert to quark pairs when traversing the QGP and the gluon degrees of freedom do not explicitly manifest in the suppression of hadron spectra at high $p_T$.

\begin{figure}[htb]
   \centering
   \epsfxsize=4.0in
   \epsfysize=3.0in
   \leavevmode
   \epsffile{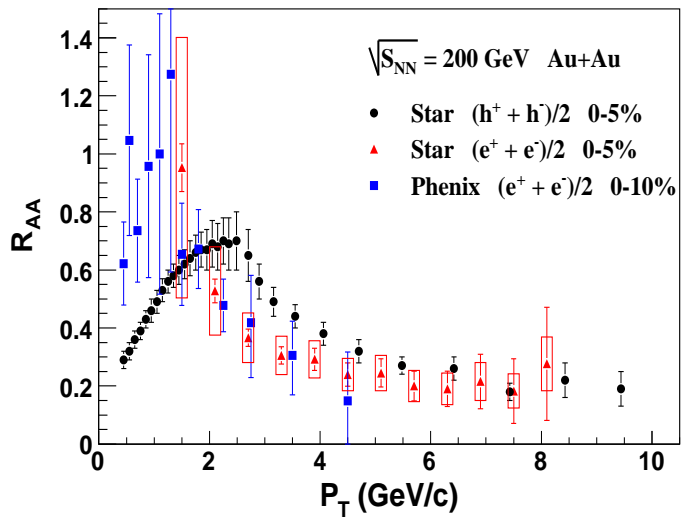} 
\parbox{5.5in}{\caption[elec-h]{Measurements of nuclear modification factor of non-photonic electrons from heavy quark semi-leptonic decays by STAR and PHENIX in comparison with that of charged hadrons by STAR. The magnitudes of suppression for hadrons and non-photonic electrons approach each other at high $P_T$.}}
\label{elec-h}
\end{figure}

Figure~\ref{elec-h} shows the STAR~\cite{star-ele} and PHENIX~\cite{phenix-ele} measurements of nuclear modification factor for non-photonic electrons in comparison with that of charged hadrons~\cite{star-h} from central Au+Au collisions at $\sqrt{s_{_{NN}}}=200$ GeV. The boxes (bars) on STAR electron data indicate the size of systematic (statistical) errors. For clarity only the statistical errors for PHENIX data are shown in the Figure and the systematic error is larger than or comparable to statistical error at each point~\cite{phenix-ele}. In the overlapping $p_{T}$ region the STAR and PHENIX data are consistent with each other within statistical and systematic errors. Non-photonic electrons in nuclear collisions are mostly from semi-leptonic decays of heavy quarks and due to decay kinematics the heavy quark meson $p_{T}$ is on average significantly higher than that of the decay electron. These measurements imply that the energy loss of heavy quarks is not significantly different from that of light quarks, in contradiction of theoretical expectations that heavy quarks would loss much less energy due to the dead-cone effect suppressing gluon radiation for heavy quarks~\cite{khar}. Recent theoretical calculations cannot explain the non-photonic electron and light hadron $R_{AA}$ measurements with a consistent dynamical picture~\cite{armesto,djor}.
The parton energy loss mechanism in dense QCD medium and the significance of elastic collisional energy loss are under re-examination now~\cite{mus,adil}. 
One of the major experimental uncertainties is the relative D and B decay contributions to the non-photonic electrons at high $p_{T}$. Lin proposed to use electron and hadron azimuthal angular correlations from D and B decays to study the relative D and B contributions~\cite{Lin}, which are being analyzed with STAR data now. The flavor dependence of parton energy loss continues to be a subject of both experimental and theoretical investigations.

\section{Hadronization of Bulk Partonic Matter}
Particle production from nucleus-nucleus collisions at RHIC has shown distinct $p_{T}$ scales: $p_{T}$ above 6 GeV/c or so where parton fragmentations become important and particle dependence in the nuclear modification factors disappears; $p_{T}$ below 2 GeV/c where hydrodynamic calculations can describe $p_{T}$ and elliptic flow $v_2$ distributions of $\pi$, Kaon, protons, $\Lambda$ and $\Xi$ particles~\cite{hydro}; and the intermediate $p_T$ region of 2-6 GeV/c where unique particle type dependence on particle yield and elliptic flow $v_2$ has been observed~\cite{star-L}. 
Experimental features in the nuclear modification factors and the elliptic flow at intermediate $p_T$ have been related to hadron formation mechanisms from bulk partonic matter through multi-parton dynamics such as quark coalescences and recombinations.  

\begin{figure}[htb]
   \centering
   \epsfxsize=4.0in
   \epsfysize=3.0in
   \leavevmode
   \epsffile{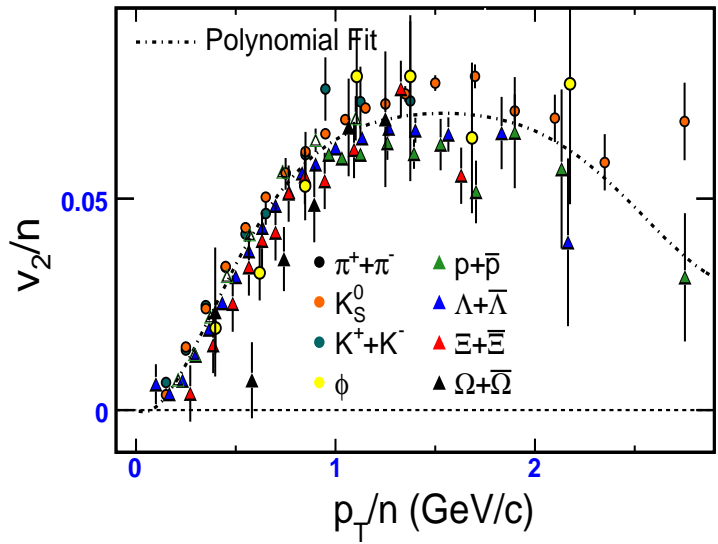} 
\parbox{5.5in}{\caption[cqm-scaling]{$v_2/n$ as a function of $p_T/n$ for identified particles, where $n$ is the number of constituent quarks. The dotted line is a polynomial fit to the available data. Systematic deviation as a function of particle masses at low $p_T$ is largely due to hydrodynamic behavior. Internal structures of hadrons also contribute to deviations.}}
\label{cqm-scaling}
\end{figure}

A unique feature of constituent quark number scaling has been observed in experimental measurements of nuclear modification factors and elliptic flow $v_2$ at intermediate $p_T$~\cite{star-L}. This scaling cannot be explained in the traditional parton fragmentation picture for hadron formation where a leading parton determines the properties of the produced hadron in nuclear collisions. Characteristics of hadrons at intermediate $p_T$ produced in nucleus-nucleus collisions at RHIC necessarily require that the constituent quarks for these hadrons pre-exist and carry an azimuthal angular anisotropy before hadronization. Figure~\ref{cqm-scaling} shows elliptic flow $v_2/n$ versus $p_T/n$, where $n$ is the constituent quark number for identified particles from the STAR and PHENIX measurements~\cite{sorensen}. At the intermediate $p_T$ region the scaled data points fall approximately on a single curve. Deviations from the scaling in the low $p_T$ region have been attributed to hydrodynamic behavior and detailed internal quark-gluon structures for various hadrons have been shown theoretically to contribute to deviations from the scaling as well~\cite{bass}. The constituent quark scaling has also been observed in the particle dependence of nuclear modification factors for $\pi$, Kaon, proton and hyperons. Recently measurements of nuclear modification factor and $v_2$ for $\phi$ mesons have exhibited the constituent quark number scaling as well~\cite{caix,sara}. The $\phi$ measurements confirm the parton origin of the dynamics for the scaling behavior and that strange quarks behave similarly to up and down quarks in nucleus-nucleus collisions at RHIC. 

Quark coalescence or recombination models~\cite{molnar,hwa,greco,fries} can explain the experimental features at the intermediate $p_T$. The constituent quark number scaling reflects the fact that hadrons are formed from constituent ingredients readily available at the hadronization stage from dense matter created in nuclear collisions and the azimuthal angular anisotropy of hadrons is the sum from individual ingredients. We denote this feature as constituent quark number scaling because the number of constituents is 2 for mesons and 3 for baryons although the constituent entity may not necessarily be the constituent quarks in traditional quark model for hadrons. 

The constituent quark number scaling in nuclear modification factors and elliptic flow for identified particles at intermediate $p_T$ originates from partonic nature of the matter. Moreover, the partonic matter must exhibit parton collectivity such as $v_2$ in order to explain experimental measurements of $v_2$ at intermediate $p_T$ for mesons and baryons including multi-strange hyperons. These data have provided most direct experimental evidence for deconfined partonic matter created in nucleus-nucleus collisions at RHIC. Experimental data seem to imply that the effective degrees of freedom near the transition temperature for dense matter created in nucleus-nucleus collisions at RHIC are in the constituent quarks. Such a picture would be qualitively consistent with the description of quark-antiquark quasi-hadron state just above the $T_c$ by Brown, Gelman and Rho~\cite{brown}. 

\section{Exotic Particle Searches at RHIC}
Experimental data at RHIC indicate that many hadrons at the intermediate $p_T$ of 2-6 GeV/c are formed from coalescences or recombinations of quarks, which are readily available from the dense partonic matter created in nuclear collisions at RHIC energies. Such a collision environment would be favorable for the formation of exotic particles if these particles do exist. Of course, searches for exotic particles from an overwhelming number of ordinary hadrons in nuclear collisions at RHIC are both scientifically and technically challenging. STAR experiment has been pursuing an active program to search for exotic particles utilizing the large acceptance Time-Projection Chamber (TPC) and its superior capability for tracking of charged particles under high multiplicity density environment. 

STAR has been carrying out experimental searches on strangelet~\cite{strangelet}, pentaquark and other exotic particles. We have observed an intriguing peak with $\sim 4\sigma$ statistical significance in $pK^{+}+\overline{p}K^{-}$ invariant mass distribution from 18.6 million d+Au collisions at $\sqrt{s_{_{NN}}}=200$ GeV. The observed peak position is at 
1.53 GeV/c$^2$ and the width is consistent with detector resolutions. We denote this peak as $\theta^{++}$ for simplicity in our description. Note the observed peak is a different charge state than what has been called $\theta^{+}$ as first reported by the LEPS collaboration~\cite{leps}. Neither the $\theta^{+}$ nor the $\theta^{++}$ states have been observed by the CLAS collaboration~\cite{clas-1,clas-2} with recent analyses of much larger statistical samples than that was previously analyzed for positive signal~\cite{clas-3}. STAR has devoted considerable efforts to understand the background and check for possible experimental bias in the analysis for $\theta^{++}$. We have also analyzed many other collision systems. We can neither confirm nor rule out the observation from d+Au collisions with currently available STAR data. Another d+Au run at RHIC will be needed to investigate the STAR data. Details of the STAR analysis can be found in reference~\cite{pentaquark}. Future STAR detector upgrades, a barrel Time-of-Flight detector under construction and a Heavy Flavor Tracker, will significantly enhance the STAR capability for exotic particle searches. 

\section{Experimental Outlook}
Searches for possible critical point in the QCD phase diagram is another interesting topic that we did not cover in this short contribution. The nature of the order of possible phase transition has not been determined experimentally. In fact, we do not know how to address the order of QCD phase transition question experimentally though generally it is believed that studies on fluctuations are relevant. Direct measurement of open charm and beauty mesons will be necessary in order to understand heavy quark dynamics in dense QCD medium, for which measurements of relative B and D contributions, elliptic flow $v_2$ of D and B mesons and elliptic flow $v_2$ for quarkonia are critical. This will require future vertex detector upgrades for both STAR and PHENIX. 

To understand the QCD properties of matter near or above the $T_c$ will require more experimental efforts. Dileptons in the low to intermediate mass regions will shed new insight on the chiral properties of the dense matter. Soft photons of tens MeV to GeV energy will be the shining lights from the dense matter and have remained an open experimental issue. And of course, our searches for exotic particles at RHIC may prove that RHIC is indeed an exotic particle factory which would allow us to study hadrons whose structure is beyond the conventional quark-antiquark and three-quark constituents. 

In summary RHIC as a dedicated QCD machine has an exciting scientific program even during the era of LHC experiment where the energy frontier is pushed an order of magnitude higher. 

\section{Acknowledgment}
We thank Xiaoyan Lin, Jingguo Ma and Paul Sorensen for providing figures and useful discussions; Chuck Whitten for careful reading of the manuscript. This paper is a contribution to a workshop dedicated to Professor Fujia Yang, former President of Fudan University and Chancellor of Nottingham University, on his 70th birthday. Professor Yang has been an inspiration for me as a scientist and an educator ever since I got to know him as a student in his Department at Fudan.

\end{document}